\documentclass[10pt]{iopart}

\usepackage{graphicx}
\usepackage{iopams}

\begin{document}

\title{The \mbox{\boldmath $t$-$t'$-$t''$-$U$} Hubbard model and Fermi-level peak}%
\author{A Sherman}%
\address{Institute of Physics, University of Tartu, W. Ostwaldi Str 1, 50411 Tartu, Estonia}%

\ead{alekseis@ut.ee}

\begin{abstract}
Using the strong coupling diagram technique, low-temperature spectral properties of the two-dimensional fermionic Hubbard model are con\-si\-de\-red for strong and moderate Hubbard repulsions $U$. The electron hopping to the nearest, second and third neighbors is taken into account with hopping constants $t$, $t'=-0.3t$ and $t''=0.2t$, respectively. The nonzero values of $t'$ and $t''$ lead to strong asymmetry in magnetic properties with respect to the hole and electron doping -- for $U=8t$ strong antiferromagnetic correlations are retained up to the electron concentration $\bar{n}\approx 1.25$, while they are destroyed completely at $\bar{n}\approx 0.87$. When the temperature is decreased to $T\lesssim 0.1t$, in a wide range of electron concentrations there appear narrow and intensive peaks at the Fermi level in densities of states. For $U\lesssim 6t$ the peaks are seen even at half-filling, while for larger $U$ they arise as the Fermi level leaves the Mott gap. The peaks are connected with a narrow band emerging at low temperatures. We identify states forming the band with spin-polaron excitations -- bound states of correlated electrons and mobile spin excitations. Obtained low-temperature spectral functions are used for interpreting the peak-dip-hump structure observed in the photoemission of Nd$_{2-x}$Ce$_x$CuO$_4$. In the case of hole doping, the calculated Fermi contour contains arcs near nodal points with pseudogaps near antinodal points, while for electron doping the spectral intensity is suppressed at hot spots, in agreement with experimental observations in cuprate perovskites.
\end{abstract}

\vspace{2pc}

%\submitto{}

\maketitle

\ioptwocol

\section{Introduction}
A realistic description of cuprate perovskites needs in a Hamiltonian with a complicated kinetic term, which contains hopping to next nearest and more distant sites (see, e.g., \cite{Brinckmann,Macridin,Das,Armitage}). These terms violate the particle-hole symmetry of the $t$-$U$ Hubbard model, which, in particular, can explain the observed significant difference in spectral and magnetic properties of $n$- and $p$-type cuprates \cite{Armitage,Fujita}. The dissimilar behavior of the antiferromagnetic (AF) ordering with doping underlies these differences. At low temperatures, in $n$-type cuprates, the AF ordering is persistent and survives up to electron concentrations $\bar{n}\approx 1.13$, adjoining or possibly overlapping with the superconducting phase \cite{Armitage}. In $p$-type cuprates, already a few percents of holes are enough to decrease the AF correlation length to several lattice spacings \cite{Damascelli}. Calculations performed in the $t$-$t'$-$t''$-$J$ model \cite{Tohyama} and in the $t$-$t'$-$t''$-$U$ Hubbard model using the cluster perturbation theory \cite{Senechal02,Senechal} also demonstrated the difference in spectral and magnetic properties of these two families of crystals. The former of these works was carried out in a small cluster, while the approach used in the latter is a method joining small cluster solutions. That is, both works imply a short AF correlation length, which is limited by sizes of used clusters. This approximation may be reasonable for doped $p$-type cuprates but is in doubt for $n$-type crystals. It would be desirable to consider both families without severe restrictions on the correlation length.

In this article, the strong-coupling diagram tech\-ni\-que (SCDT) \cite{Vladimir,Metzner,Pairault,Sherman06,Sherman18} is used for this purpose. The method is based on the series expansion of Green's functions in powers of the kinetic energy. Building blocks of the diagram technique are renormalized hopping lines and cumulants of different orders constructed from electron operators. The lin\-ked-cluster theorem is valid and partial summations are allowed in the SCDT. In the approximation used in \cite{Sherman18}, this allowed us to sum infinite sequences of ladder diagrams describing interactions of electrons with charge and spin fluctuations of all ranges. Hence the calculations can be performed for an infinite crys\-tal and no limitation on the AF correlation length is imposed. In comparison with approaches based on the dynamic mean-field approximation (DMFA) \cite{Georges} -- the cellular DMFA \cite{Maier,Park,Sato}, dynamic clu\-s\-ter approximation \cite{Maier,Moukouri,Merino14}, dynamic vertex approximation \cite{Toschi,Schafer,Rohringer} and dual fermion approach \cite{Rohringer,Rubtsov08,Hafermann} -- the SCDT calculations are simpler, no peculiarities of the Anderson impurity model (AIM) \cite{Hewson} are introduced from the very beginning and no special remedy to correct the overestimated Mott critical repulsion is necessary. As for the clu\-s\-ter perturbation theory, the variation cluster ap\-p\-ro\-xi\-ma\-ti\-on \cite{Potthoff,Arrigoni,Faye} imposes a constraint on the range of AF fluctuations, which, as mentioned, can be avoided in SCDT. Conceptually, the approach is close to the diagram technique for Hubbard operators \cite{Zaitsev,Izyumov88,Izyumov90,Ovchinnikov}. However, the number of diagrams describing the same processes is much smaller in SCDT. An additional difficulty of the technique for Hubbard operators is the dependence of its rules and graphic representation on the choice of the operator precedence. In the SCDT this problem is absent. In \cite{Sherman18,Sherman18a}, it was shown that for the $t$-$U$ Hubbard model the spectral function, density of states (DOS), zero-frequency uniform susceptibility, double occupancy and squared site spin calculated using the SCDT are in reasonable agreement with results of the exact diagonalization, Monte Carlo simulations and some of the above-mentioned approaches in wide ranges of temperatures, Hubbard repulsions, and doping.

In this work, the SCDT is used for calculating spectral functions, densities of states, zero-frequency staggered magnetic susceptibilities $\chi({\bf Q},\omega=0)$ and Fermi contours in the $t$-$t'$-$t''$-$U$ Hubbard model with $t'=-0.3t$ and $t''=0.2t$. These parameters are suggested by band-structure calculations \cite{Andersen} and are close to those used in other works on the electron structure of cuprates \cite{Brinckmann,Macridin,Das,Armitage,Tohyama,Senechal}. Here ${\bf Q}=(\pi,\pi)$ is the AF ordering vector (hereinafter the intersite distance is set as the unit of length). Since cuprates are crystals with strong electron correlations, the calculations were mainly carried out for values of the Hubbard repulsion in this range -- $U=5.1t$ and $U=8t$. In agreement with experiment \cite{Armitage,Fujita,Damascelli} and earlier calculations \cite{Tohyama,Senechal}, our results show a strong doping asymmetry of the AF ordering and spectral functions in this model. In particular, for $U=8t$ and the temperature $T=0.13t$ the quantity $\chi({\bf Q},0)$, which characterizes the proximity to the long-range AF order, is large and varies only weakly in the range $1\leq\bar{n}\lesssim 1.25$, while it is small for hole doping at $\bar{n}\approx 0.87$, signaling the destruction of AF correlations. For $U=5.1t$, $T=0.12t$ and the underdoped case, the Fermi contour consists of Fermi arcs near the momenta $(\pm\pi/2,\pm\pi/2)$, $(\pm\pi/2,\mp\pi/2)$ with pseudogaps around $(\pm\pi,0)$, $(0,\pm\pi)$ for hole doping, while the contour is diamond-shaped with intensity suppressions at hot spots for electron doping. Similar contour shapes are observed in photoemission of cuprates \cite{Armitage,Damascelli}.

Considering the $t$-$U$ Hubbard model in \cite{Sherman18a}, we found that, as the temperature decreased, a narrow peak appeared in the DOS on the Fermi level (FL). For the repulsion $U=5.1t$, which is smaller than the critical value of the Mott transition $U_c\approx 6t$, the peak existed even at half-filling in a DOS dip inherent in the bad-metal state at higher temperatures. The peak persisted at the FL in some doping range. It was found that the peak originated from a band, which arose at a low $T$ and for half-filling had a width of the order of the exchange constant $J=4t^2/U$. This fact suggested qualifying the band as the spin-polaron one, which consists of bound states of electrons and spin excitations -- the notion borrowed from the $t$-$J$ model \cite{Schmitt,Ramsak,Sherman94}. For the appearance of these states, the spin-excitation branch has to be well formed that implies low temperatures and large correlation lengths. Indeed, in the present work, we found that the peak and the band are visible when there is an intensive maximum of the zero-frequency susceptibility at {\bf Q}. In the $t$-$t'$-$t''$-$U$ model the range of $\bar{n}$, in which these spectral peculiarities reveal themselves, is much wider than in the $t$-$U$ model. This allows us to consider their concentration evolution in more detail. Besides, in this work, the appearance of the low-temperature peculiarities is considered for larger $U$. For $U>U_c$, the peak arises once the FL leaves the Mott gap. Hence the peak has some similarity with the quasiparticle peak of the DMFA \cite{Georges} -- at half-filling, both peaks appear in the DOS when $U$ becomes smaller than $U_c$, both peaks are manifestations of bound states of electrons and spin excitations. However, if in the DMFA the peak is a modified Kondo or Abrikosov-Suhl resonance, which is related to bound states of {\sl free} electrons and {\sl localized} spins \cite{Hewson}, in the SCDT we are dealing with bound states of {\sl correlated} electrons and {\sl mobile} spin excitations. In the latter approach, the flatness of regions in the spin-polaron band, which are responsible for the peak, is not connected with the locality of the underlying reference system, as in the DMFA. Rather it is a consequence of the fact that the band is pinned to the FL, which leads to its flattening with doping. A similar flattening of the spin-polaron band and a persistent DOS peak at the FL were observed in the $t$-$J$ model \cite{Sherman94,Sherman97}. In two dimensions (2D), the momentum dependence of the electron self-energy is rather strong \cite{Maier,Rohringer,Sherman17}, and the applicability of the DMFA ideas, in this case, is not evident. We argue that the peak at the FL (for brevity, the FL peak) may explain some peculiarities of photoemission spectra in the $n$-type crystal Nd$_{2-x}$Ce$_x$CuO$_4$ \cite{Armitage01,Armitage02,Matsui05,Matsui}.

\section{Main formulas}
Since main equations used in this work were deduced in detail in \cite{Sherman18,Sherman18a}, here they are given without derivation, for clarity sake of the following discussion.

The Hamiltonian of the 2D fermionic Hubbard model \cite{Hubbard} reads
\begin{equation}\label{Hamiltonian}
H=\sum_{\bf ll'\sigma}t_{\bf ll'}a^\dagger_{\bf l'\sigma}a_{\bf l\sigma}
+\frac{U}{2}\sum_{\bf l\sigma}n_{\bf l\sigma}n_{\bf l,-\sigma},
\end{equation}
where vectors ${\bf l}$ and ${\bf l'}$ label sites of a square plane lattice, $\sigma=\uparrow,\downarrow$ is the spin projection, $a^\dagger_{\bf l\sigma}$ and $a_{\bf l\sigma}$ are electron creation and annihilation operators, $t_{\bf ll'}$ are  hopping constants and $n_{\bf l\sigma}=a^\dagger_{\bf l\sigma}a_{\bf l\sigma}$. As mentioned in the Introduction, in this work, hopping constants between the nearest $t$, second $t'=-0.3t$ and third $t''=0.2t$ neighbors are taken into account.

The following one- and two-particle Green's fun\-c\-ti\-ons
\begin{eqnarray}
G({\bf l',\tau';l,\tau})&=&\langle{\cal T}\bar{a}_{\bf l'\sigma}(\tau')
a_{\bf l\sigma}(\tau)\rangle,\label{Glt}\\
\chi({\bf l',\tau';l,\tau})&=&\langle{\cal T}s^+_{\bf l'}(\tau')
s^-_{\bf l}(\tau)\rangle\nonumber\\
&=&\langle{\cal T}\bar{a}_{\bf l'\uparrow}(\tau')a_{\bf l'\downarrow}(\tau')\bar{a}_{\bf l\downarrow}(\tau)a_{\bf l\uparrow}(\tau)\rangle\label{clt}
\end{eqnarray}
are considered. Here the statistical averaging denoted by the angular brackets and time dependencies $$\bar{a}_{\bf l\sigma}(\tau)=\exp{({\cal H}\tau)}a^\dagger_{\bf l\sigma}\exp{(-{\cal H}\tau)}$$
are determined by the operator ${\cal H}=H-\mu\sum_{\bf l\sigma}n_{\bf l\sigma}$ with the chemical potential $\mu$. The time-ordering operator ${\cal T}$ arranges operators from right to left in ascending order of times $\tau$. Up to the constant factor $\chi({\bf l',\tau';l,\tau})$ (\ref{clt}) coincides with the spin susceptibility.

\begin{figure}[t]
\centerline{\resizebox{0.95\columnwidth}{!}{\includegraphics{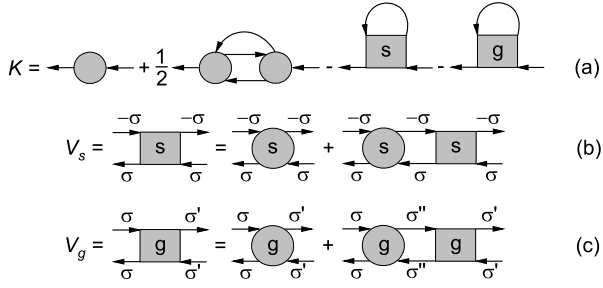}}}
\caption{(a) Diagrams in $K({\bf k},j)$ taken into account in the present work. (b,c) Bethe-Salpeter equations for the four-leg diagrams in part (a).} \label{Fig1}
\end{figure}

For calculating these quantities the SCDT \cite{Vladimir,Metzner,Pairault,Sherman06} is used (a concise description of this diagram technique can be found in \cite{Sherman17}). In this approach, the Fourier transform of Green's function (\ref{Glt}) can be presented in the form
\begin{equation}\label{Larkin}
G({\bf k},j)=\Big\{\big[K({\bf k},j)\big]^{-1}-t_{\bf k}\Big\}^{-1},
\end{equation}
where ${\bf k}$ is the wave vector, $j$ is an integer defining the Matsubara frequency $\omega_j=(2j-1)\pi T$, $t_{\bf k}$ is the Fourier transform of $t_{\bf ll'}$ and $K({\bf k},j)$ is the irreducible part -- the sum of all two-leg irreducible diagrams, which cannot be divided into two disconnected parts by cutting a hopping line. Diagrams taken into account in this work for calculating $K({\bf k},j)$ are shown in figure \ref{Fig1}(a).
Here internal arrowed lines are the renormalized hopping terms
\begin{equation}\label{hopping}
\theta({\bf k},j)=t_{\bf k}+t^2_{\bf k}G({\bf k},j),
\end{equation}
and circles without letters are on-cite cumulants of the first and second orders
\begin{eqnarray*}
&&C_1(\tau',\tau)=\big\langle{\cal T}\bar{a}_{{\bf l}\sigma}(\tau')a_{{\bf l}\sigma}(\tau)\big\rangle_0,\\
&&C_2(\tau_1,\sigma_1;\tau_2,\sigma_2;\tau_3,\sigma_3;\tau_4,\sigma_4)\\
&&\quad\quad=\big\langle{\cal T}\bar{a}_{{\bf l}\sigma_1}(\tau_1)a_{{\bf l}\sigma_2}(\tau_2) \bar{a}_{{\bf l}\sigma_3}(\tau_3)a_{{\bf l}\sigma_4}(\tau_4)\big\rangle_0\\
&&\quad\quad-\big\langle{\cal T}\bar{a}_{{\bf l}\sigma_1}(\tau_1)a_{{\bf l}\sigma_2}(\tau_2)\big\rangle\big\langle{\cal T}\bar{a}_{{\bf l}\sigma_3}(\tau_3)a_{{\bf l}\sigma_4}(\tau_4)\big\rangle_0\\
&&\quad\quad+\big\langle{\cal T}\bar{a}_{{\bf l}\sigma_1}(\tau_1)a_{{\bf l}\sigma_4}(\tau_4)\big\rangle\big\langle{\cal T}\bar{a}_{{\bf l}\sigma_3}(\tau_3)a_{{\bf l}\sigma_2}(\tau_2)\big\rangle_0,
\end{eqnarray*}
where the subscript 0 near brackets indicates that time dependencies and averages are determined by the local operator ${\cal H}_{\bf l}=\sum_\sigma\left[(U/2)n_{\bf l\sigma}n_{\bf l,-\sigma}- \mu n_{\bf l\sigma}\right]$. The last two terms in $K({\bf k},j)$ correspond to diagrams with ladder inserts. The vertices $V_s$ and $V_g$ satisfy the Bethe-Salpeter equations shown in figures~\ref{Fig1}(b) and (c), in which circles with letters are irreducible four-leg diagrams which cannot be divided into two disconnected parts by cutting a pair of horizontal oppositely directed hopping lines. In this work, the\-se irreducible diagrams are approximated by the respective second-order cumulants. In this case, $V_s$ and $V_g$, apart from frequencies, depend only on the transfer momentum. As a consequence, the irreducible part reads
\begin{eqnarray}\label{K}
&&K({\bf k},j)=C_1(j)\nonumber\\
&&\quad\quad-\frac{T}{N}\sum_{{\bf k'}j'}\theta({\bf k'},j')\big[V_{s,\bf k- k'}(j,\sigma;j,\sigma;j',-\sigma;j',-\sigma)\nonumber\\
&&\quad\quad+V_{g,\bf k-k'}(j,\sigma;j,\sigma;j',\sigma;j',\sigma)\big]\nonumber\\
&&\quad\quad+\frac{T^2}{2N^2}\sum_{{\bf k'}j'\nu}\theta({\bf k'},j'){\cal T}_{\bf k-k'}(j+\nu,j'+\nu) \nonumber\\
&&\quad\quad\times\Big[C_2(j,\sigma;j+\nu,\sigma;j'+\nu,-\sigma;j',-\sigma)\nonumber\\
&&\quad\quad\times C_2(j+\nu,\sigma;j,\sigma;j',-\sigma;j'+\nu,-\sigma)\nonumber\\
&&\quad\quad+\sum_{\sigma'} C_2(j,\sigma;j+\nu,\sigma';j'+\nu,\sigma';j',\sigma)\nonumber\\
&&\quad\quad\times C_2(j+\nu,\sigma';j,\sigma;j',\sigma;j'+\nu,\sigma')\Big].
\end{eqnarray}

The vertex $V_s$ is connected with the spin su\-s\-cep\-ti\-bi\-li\-ty (\ref{clt}) by the relation
\begin{eqnarray}\label{chi}
\chi({\bf k},\nu)&=&-\frac{T}{N}\sum_{{\bf q}j}G({\bf q},j)G({\bf q+k},j+\nu) \nonumber\\
&&-T^2\sum_{jj'}F_{\bf k}(j,j+\nu)F_{\bf k}(j',j'+\nu)\nonumber\\
&&\times V_{s\bf k}(j+\nu,\sigma;j'+\nu,\sigma;j',-\sigma;j,-\sigma),
\end{eqnarray}
where
\begin{eqnarray*}
&&F_{\bf k}(j,j')=N^{-1}\sum_{\bf q}\Pi({\bf q},j)\Pi({\bf q+k},j'),\\
&&\Pi({\bf k},j)=1+t_{\bf k}G({\bf k},j),
\end{eqnarray*}
$N$ is the number of sites and $\nu$ defines the boson Matsubara frequency $\omega_\nu=2\nu\pi T$. It can be shown that $V_s$ coincides with the antisymmetrized part of $V_g$, $\sigma\sum_{\sigma'}\sigma'V_g(\sigma',\sigma,\sigma,\sigma')$, while the symmetrized part $V_c=\sum_{\sigma'}V_g(\sigma',\sigma,\sigma,\sigma')$ is connected with the charge susceptibility by the relation similar to (\ref{chi}).

The Bethe-Salpeter equations for $V_s$ and $V_c$ are significantly simplified in the range of $\mu$ defined by the inequalities
\begin{equation}\label{condition}
T\ll\mu,\quad T\ll U-\mu.
\end{equation}
For $U\gg T$ this range contains the most interesting cases of half-filling and moderate doping. For these $\mu$, the Bethe-Salpeter equations in figures~\ref{Fig1}(b) and (c) can be written as follows:
\begin{eqnarray}\label{Vs}
&&V_{s\bf k}(j+\nu,j,j',j'+\nu)=\frac{1}{2}f_{1\bf k}(j+\nu,j'+\nu)\nonumber\\
&&\quad\quad\times\bigg\{\bigg[a_2(j'+\nu,j+\nu)-\frac{\delta_{jj'}}{T+\zeta} a_1(j'+\nu)\bigg] \nonumber\\
&&\quad\quad\times\big[a_1(j)+y_{1\bf k}(j,j')\big]-\frac{\delta_{\nu0}}{2(T+\zeta)}a_1(j)a_1(j') \nonumber\\
&&\quad\quad+a_1(j'+\nu)\big[a_2(j,j')+y_{2\bf k}(j,j')\big]\nonumber\\
&&\quad\quad+a_3(j'+\nu,j+\nu)\big[a_4(j,j')+y_{4\bf k}(j,j')\big]\nonumber\\
&&\quad\quad+a_4(j'+\nu,j+\nu)\big[a_3(j,j')+y_{3\bf k}(j,j')\big]\bigg\},
\end{eqnarray}
\begin{eqnarray}\label{Vc}
&&V_{c\bf k}(j+\nu,j,j',j'+\nu)=-\frac{1}{2}f_{2\bf k}(j+\nu,j'+\nu)\nonumber\\
&&\quad\quad\times\bigg\{a_2(j'+\nu,j+\nu)\big[a_1(j)+z_{1\bf k}(j,j')\big]\nonumber\\
&&\quad\quad-\frac{3\delta_{\nu0}}{2(T+\zeta)}a_1(j)a_1(j') \nonumber\\
&&\quad\quad+a_1(j'+\nu)\big[a_2(j,j')+z_{2\bf k}(j,j')\big]\nonumber\\
&&\quad\quad+a_3(j'+\nu,j+\nu)\big[a_4(j,j')+z_{4\bf k}(j,j')\big]\nonumber\\
&&\quad\quad+a_4(j'+\nu,j+\nu)\big[a_3(j,j')+z_{3\bf k}(j,j')\big]\bigg\},
\end{eqnarray}
where
\begin{eqnarray*}
&&f_{1\bf k}(j,j')=\bigg[1+\frac{1}{4}a_1(j)a_1(j'){\cal T}_{\bf k}(j,j')\bigg]^{-1},\\
&&f_{2\bf k}(j,j')=\bigg[1-\frac{3}{4}a_1(j)a_1(j'){\cal T}_{\bf k}(j,j')\bigg]^{-1},\\
&&{\cal T}_{\bf k}(j,j')=N^{-1}\sum_{\bf k'}\theta({\bf k+k'},j)\theta({\bf k'},j'),\\
&&g_1(j)=(i\omega_j+\mu)^{-1},\quad g_2(j)=(i\omega_j+\mu-U)^{-1}, \\
&&a_1(j)=g_1(j)-g_2(j),\quad a_2(j,j')=g_1(j)g_1(j'),\\
&&a_3(j,j')=g_2(j)-g_1(j'),\quad a_4(j,j')=a_1(j)g_2(j'),
\end{eqnarray*}
and quantities $y_{i\bf k}(j,j')$ and $z_{i\bf k}(j,j')$, $i=1,\ldots 4$ are solutions of two systems of four linear equations
\begin{eqnarray}\label{ykj}
&&y_{i\bf k}(j,j')=b_{i\bf k}(j,j')+\bigg[c_{i2}({\bf k},j-j')- \frac{\delta_{jj'}}{T+\zeta}\nonumber\\
&&\;\times c_{i1}({\bf k},j-j')\bigg] y_{1\bf k}(j,j')+c_{i1}({\bf k},j-j')y_{2\bf k}(j,j')\nonumber\\
&&\;+c_{i4}({\bf k},j-j')y_{3\bf k}(j,j')+c_{i3}({\bf k},j-j')y_{4\bf k}(j,j'),
\end{eqnarray}
\begin{eqnarray}\label{zkj}
&&z_{i\bf k}(j,j')=d_{i\bf k}(j,j')-e_{i2}({\bf k},j-j')z_{1\bf k}(j,j')\nonumber\\
&&\;-e_{i1}({\bf k},j-j')z_{2\bf k}(j,j')-e_{i4}({\bf k},j-j')z_{3\bf k}(j,j')\nonumber\\
&&\;-e_{i3}({\bf k},j-j')z_{4\bf k}(j,j').
\end{eqnarray}
Thus, the solution of the two Bethe-Salpeter equations was reduced to the solution of two small systems of linear equations (\ref{ykj}) and (\ref{zkj}). The equations depend parametrically on {\bf k}, $j$ and $j'$. In these equations,
\begin{eqnarray*}
&&b_{i\bf k}(j,j')=-\frac{1}{4}a_i(j,j')a_1(j)a_1(j'){\cal T}_{\bf k}(j,j')f_{1\bf k}(j,j')\\
&&\quad+\bigg[a_2(j,j')-\frac{\delta_{jj'}}{T+\zeta}a_1(j)\bigg]c_{i1}({\bf k},j-j')\\
&&\quad+a_1(j)c_{i2}({\bf k},j-j')+a_4(j,j')c_{i3}({\bf k},j-j')\\
&&\quad+a_3(j,j')c_{i4}({\bf k},j-j'),\\
&&d_{i\bf k}(j,j')=\frac{3}{4}a_i(j,j')a_1(j)a_1(j'){\cal T}_{\bf k}(j,j')f_{2\bf k}(j,j')\\
&&\quad-a_2(j,j')e_{i1}({\bf k},j-j')-a_1(j)e_{i2}({\bf k},j-j')\\
&&\quad-a_4(j,j')e_{i3}({\bf k},j-j')-a_3(j,j')e_{i4}({\bf k},j-j'),\\
&&c_{ii'}({\bf k},\nu)=\frac{T}{2}\sum_j a_i(j+\nu,j)a_{i'}(j,j+\nu)\\
&&\quad\times{\cal T}_{\bf k}(j+\nu,j)f_{1\bf k}(j+\nu,j),\\
&&e_{ii'}({\bf k},\nu)=\frac{T}{2}\sum_j a_i(j+\nu,j)a_{i'}(j,j+\nu)\\
&&\quad\times{\cal T}_{\bf k}(j+\nu,j)f_{2\bf k}(j+\nu,j).
\end{eqnarray*}

Expressions (\ref{Larkin}), (\ref{K}), (\ref{Vs})--(\ref{zkj}) form a closed set of equations for calculating the one-particle Green's function (\ref{Glt}). This set can be solved by iteration. As a starting function in this iteration, the Hubbard-I solution \cite{Hubbard} was used. This solution is obtained from the above formulas if the irreducible part is approximated by $C_1$ -- the first term in the right-hand side of (\ref{K}) \cite{Vladimir}. No artificial broadening was introduced in these calculations. Derived one-particle Green's functions and vertices were subsequently used for calculating the spin susceptibility.

\begin{figure*}[htb]
\centerline{\resizebox{1.99\columnwidth}{!}{\includegraphics{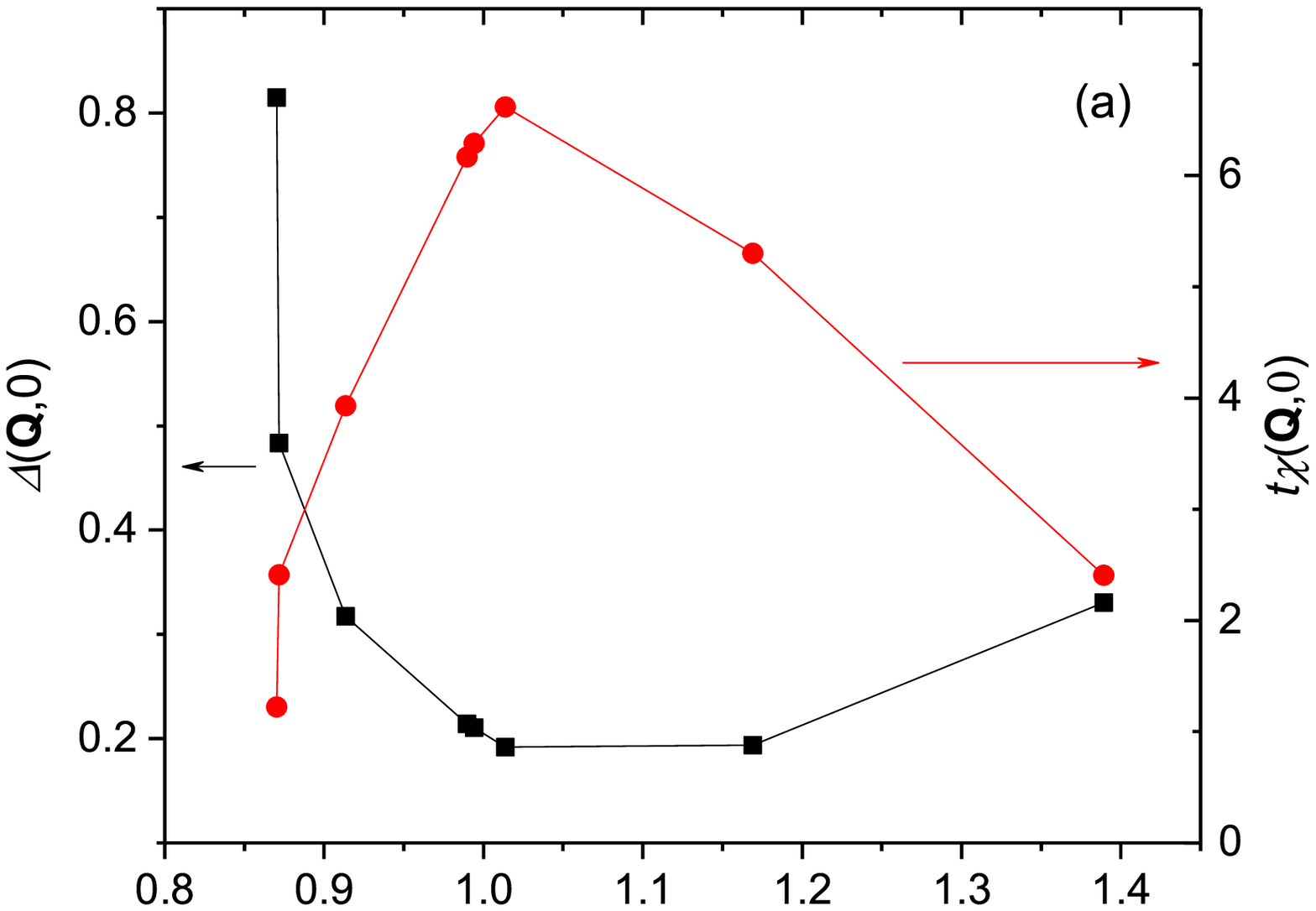}\hspace{3em}
\includegraphics{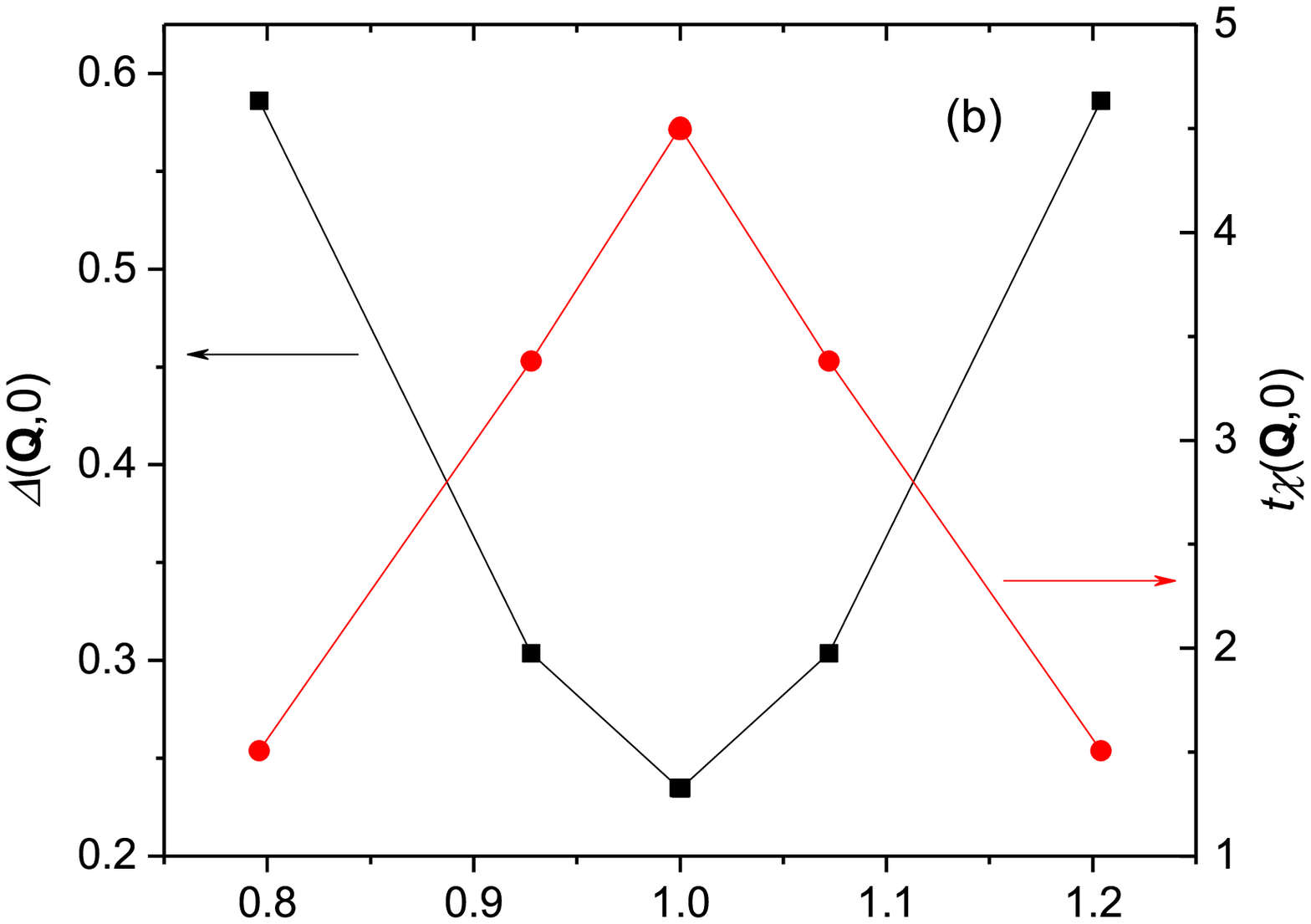}}}
\vspace{3ex}
\centerline{\resizebox{1.99\columnwidth}{!}{\includegraphics{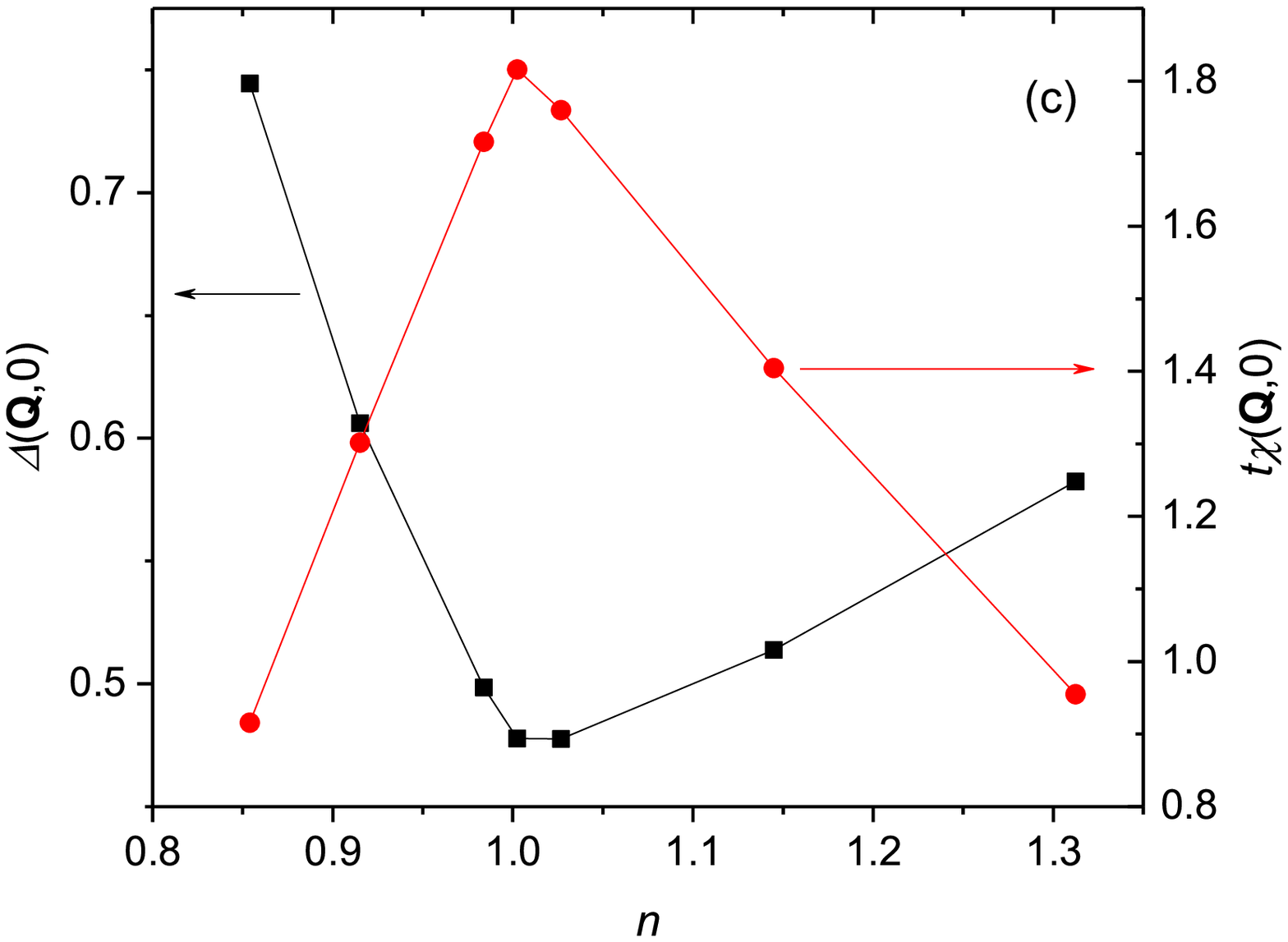}\hspace{3em}
\includegraphics{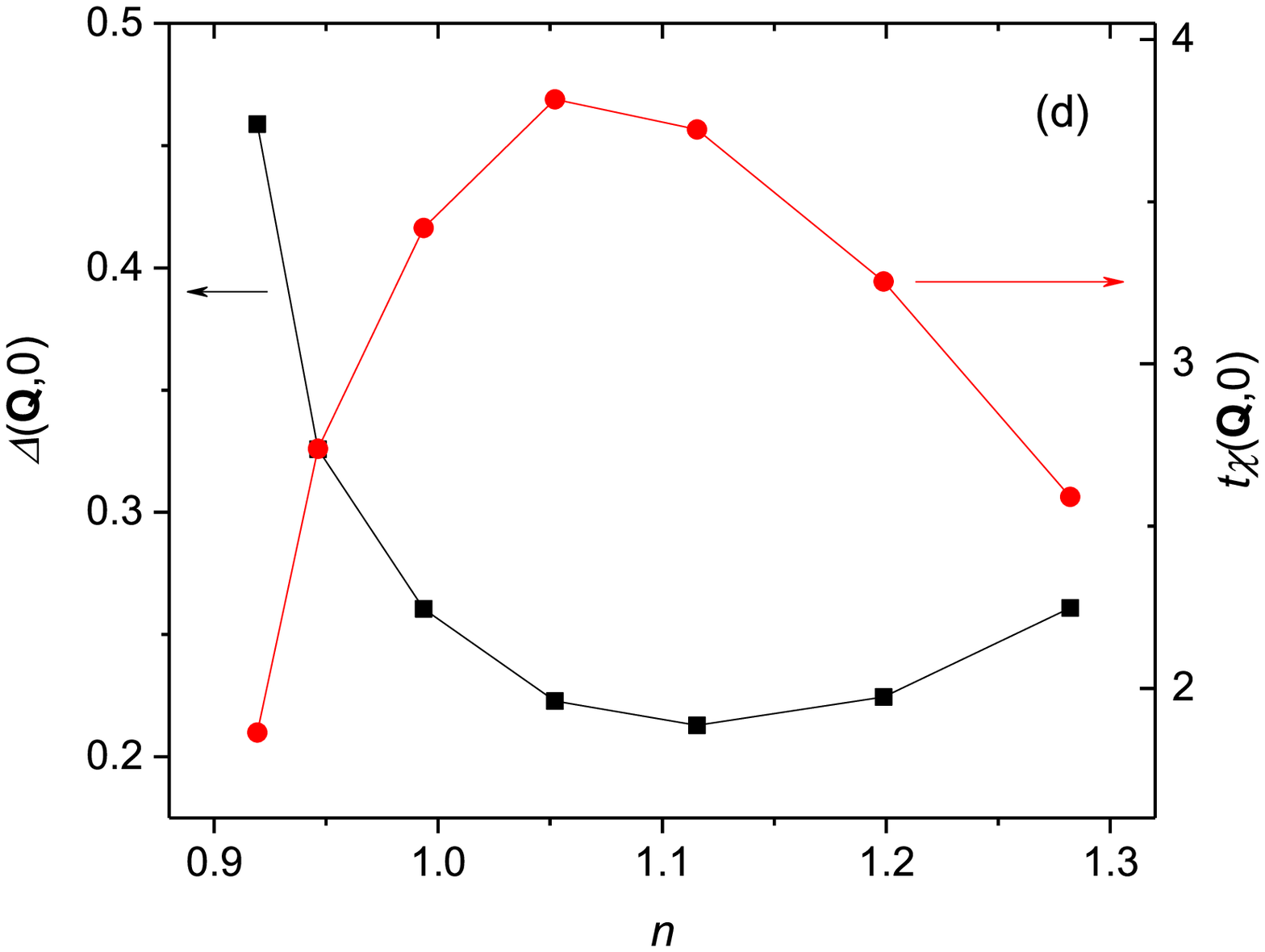}}}
\caption{Concentration dependencies of the determinant $\Delta({\bf Q},0)$ of the system \protect(\ref{ykj}) (black squares) and the zero-frequency magnetic susceptibility $\chi({\bf Q},0)$ \protect(\ref{chi}) (red circles). Panels (a), (c) and (d) correspond to the $t$-$t'$-$t''$-$U$ model with parameters $U=8t$, $T=0.13t$ (a), $U=8t$, $T=0.32t$ (c) and $U=5.1t$, $T=0.12t$ (d). Panel (b) shows results for the $t$-$U$ model with $U=8t$ and $T=0.13t$.} \label{Fig2}
\end{figure*}
It is worth noting that the above equations cor\-r\-es\-pond to an infinite crystal since infinite sequences of ladder diagrams were summed without any limitations on site indices. Finite meshes of {\bf k}-points used in calculations correspond to the method of the numerical integration rather than to a finite size of the lattice. These infinite sequences of diagrams allow us to take into account charge and spin fluctuations of all ranges.

The above equations describe the transition to the long-range AF order. The transition occurs when the determinant $\Delta({\bf k},j-j')$ of the system of linear equations (\ref{ykj}) vanishes. For all considered sets of parameters, this happens first at $j=j'$ and ${\bf k}={\bf Q}$. The vanishing $\Delta$ leads to the divergence of $y_i$ in (\ref{ykj}), which entails the divergence of the vertex $V_s$ (\ref{Vs}) and the spin susceptibility (\ref{clt}). Notice that for all considered parameters no indication of a divergence in the charge vertex and susceptibility was found. We interpret this result as the lack of a charge ordering in the model. With the parameter $\zeta=0$ in the above equations, the transition to the long-range AF order occurs at a small but fi\-ni\-te temperature $T_{\rm AF}$. This result is inconsistent with the Mermin-Wagner theorem \cite{Mermin}. In \cite{Sherman18a}, the parameter $\zeta$ was introduced in the second-order cumulant to remedy this defect. This amendment is determined by cumulants of higher orders, which form corrections to the used irreducible four-leg vertex -- the second-order cumulant. Complicated expressions of these corrections do not allow us to calculate $\zeta$ from the serial expansion. Instead, in \cite{Sherman18a} and in this work, it is determined from the condition $T_{\rm AF}=0$ for $\mu=U/2$, which reduces to $\Delta({\bf k}={\bf Q},j-j'=0)=0$ for this chemical potential and $T=0$. To check this approximation some magnetic parameters, double occupancy and squared local spin were calculated for the $t$-$U$ model at low temperatures in \cite{Sherman18a}. These parameters, their temperature and concentration dependencies appeared to be in satisfactory agreement with Monte Carlo data. In connection with the present work, it should be noted that in the $t$-$U$ model the case $\mu=U/2$ corresponds to half-filling and to the smallest value of $\Delta({\bf Q},0)$ at given parameters. As will be seen below, in the $t$-$t'$-$t''$-$U$ model, $\mu=U/2$ corresponds to half-filling only approximately for large repulsions and does not provide the lowest value to $\Delta({\bf Q},0)$. However, for all considered $U$, this value is close to the smallest one. Therefore, in this work, $\zeta$ was also determined at $\mu=U/2$. For the cases $U=8t$ and $5.1t$ considered below, $\zeta=0.145t$, which is smaller than $0.24t$ in the $t$-$U$ model with these repulsions due to frustrations introduced by extra hopping terms.

\section{Results and discussion}
\subsection{The particle-hole asymmetry}
As follows from the above discussion, small values of the determinant $\Delta({\bf Q},0)$ can serve as an indicator of the proximity to the long-range AF order. Another indicator of this proximity is a large value of the zero-frequency susceptibility at the AF momentum $\chi({\bf Q},0)$. Both these quantities are shown in Fig.~\ref{Fig2}(a) as functions of the electron concentration $\bar{n}=2\langle a^\dagger_{\bf l\sigma}a_{\bf l\sigma}\rangle$ for the case of a large repulsion and a low temperature. Both curves are strongly asymmetric with respect to the point of half-filling, $\bar{n}=1$ -- both $\Delta({\bf Q},0)$ and $\chi({\bf Q},0)$ vary only slightly in the range $1\leq\bar{n}\lesssim 1.25$, while their dependencies are very strong on the hole-doping side. For the used parameters the short-range AF ordering remains practically unchanged in the mentioned large range of electron doping, and it is completely destroyed by 13\% of holes. This destruction manifests itself in nearly discontinuous variations of the two quantities near this concentration. A similar breakdown of AF correlations was observed in the $t$-$J$ model \cite{Sherman97}. The difference in the strength of the correlations in cases of the hole and electron doping resemble that observed in cuprates \cite{Armitage,Fujita} and in calculations on small clusters \cite{Tohyama}. In real crystals, this difference is further enhanced by an interlayer spin interaction, which in the case of $n$-type cuprates increases significantly the correlation length and may apparently stabilize the long-range AF ordering up to electron concentrations of the superconducting pha\-se \cite{Armitage}.  For comparison, Fig.~\ref{Fig2}(b) demonstrates concentration dependencies of the same quantities in the $t$-$U$ model with the same $U$ and $T$. Neither the rigidity of the AF ordering for electron doping nor its extreme sensitivity for hole doping is seen in this case. In the $t$-$t'$-$t''$-$U$ model, some asymmetry of the influence of electron and hole doping on the AF ordering is retained with increasing temperature (see Fig.~\ref{Fig2}(c)). However, this asymmetry is less pronounced in comparison with lower $T$.

\begin{figure}[tbh]
\resizebox{0.95\columnwidth}{!}{\includegraphics{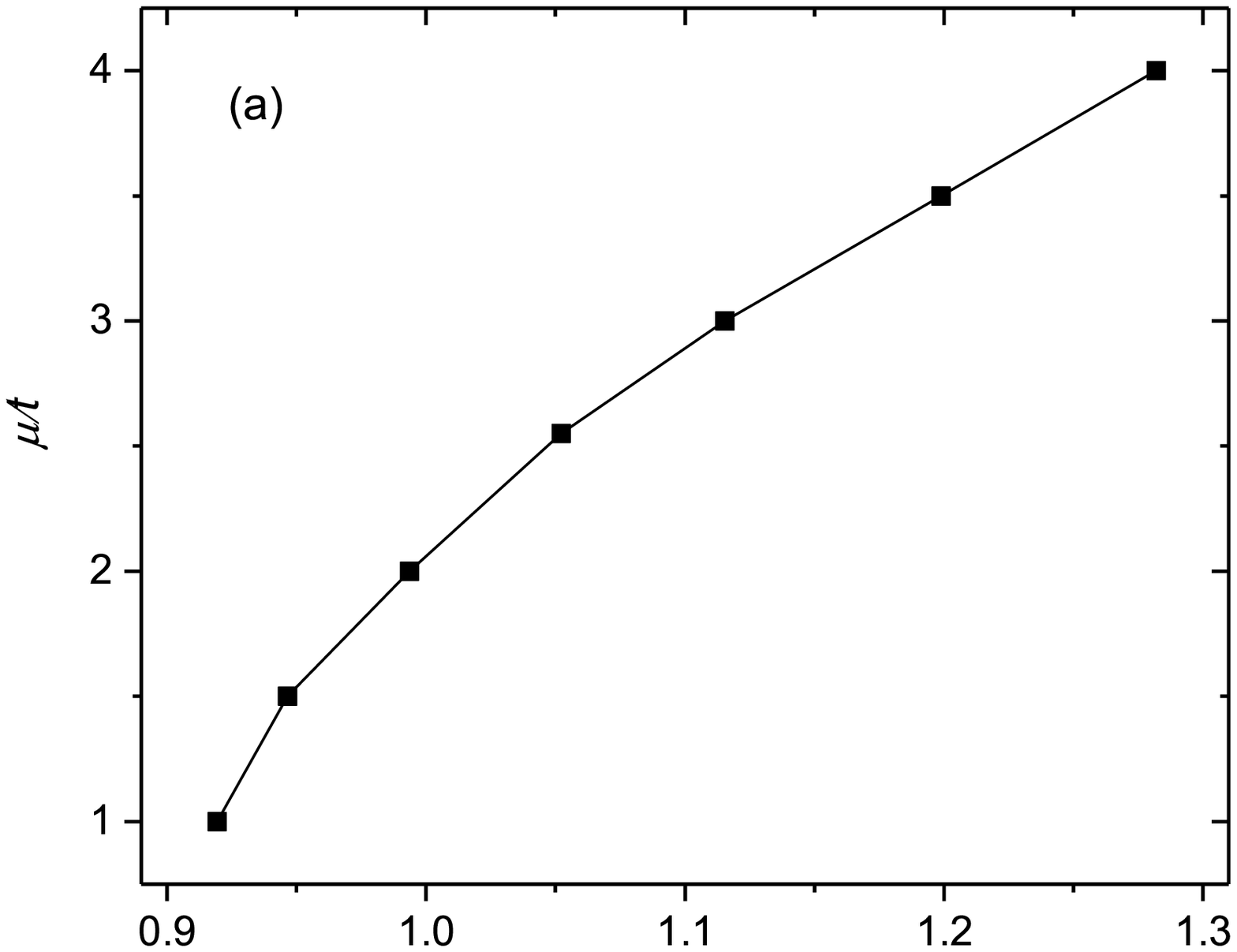}}\vspace{2ex}
\resizebox{0.95\columnwidth}{!}{\includegraphics{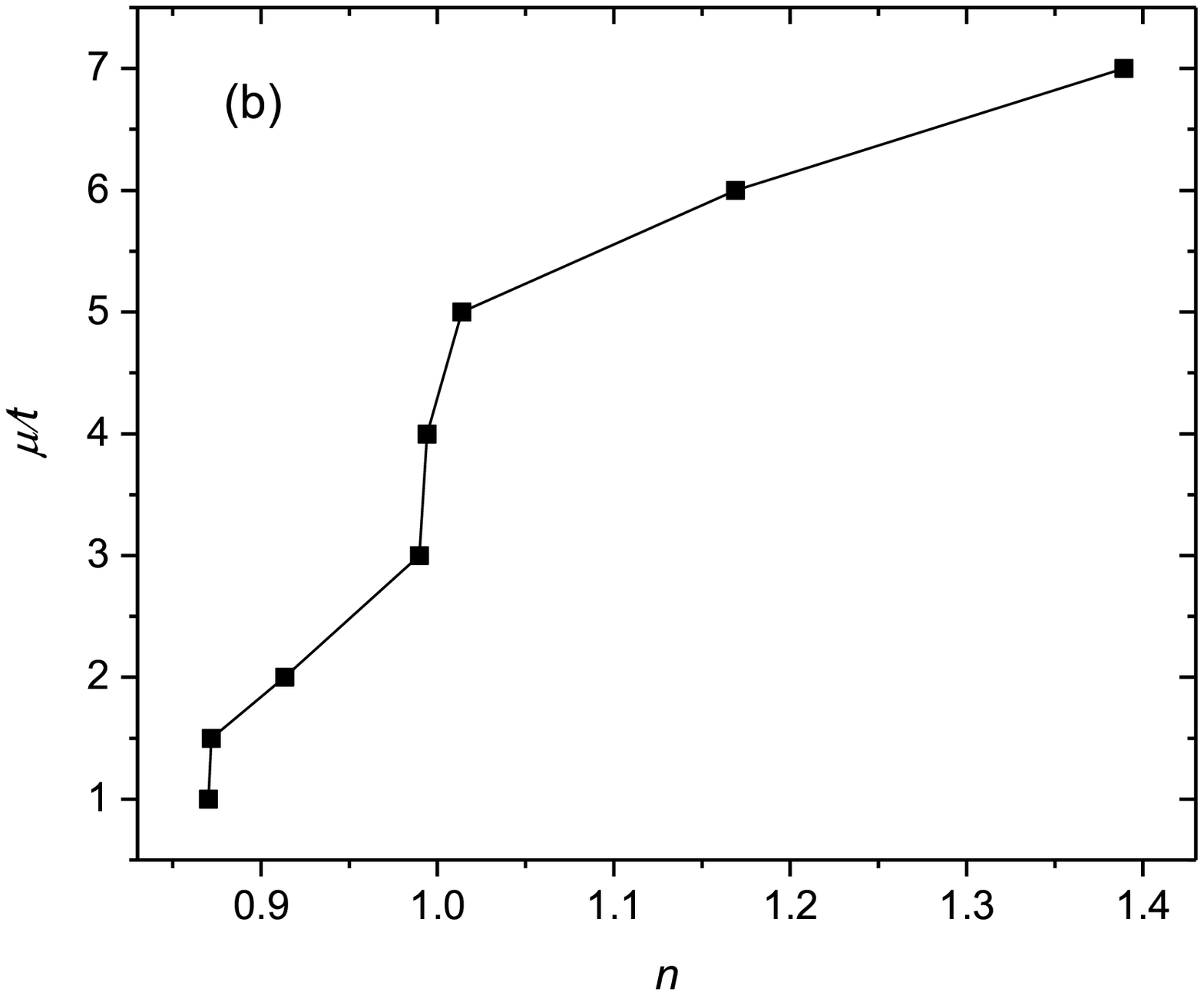}}
\caption{The dependence $\mu$ vs $\bar{n}$ in the $t$-$t'$-$t''$-$U$ Hubbard mo\-del for $U=5.1t$, $T=0.12t$ (a) and $U=8t$, $T=0.13t$ (b).} \label{Fig3}
\end{figure}
The doping asymmetry is retained also for a smaller $U$. This case is shown in Fig.~\ref{Fig2}(d). There are no apparent indications of a sudden destruction of the AF order here. For both panels (a) and (d) calculations were carried out in the range $t\leq\mu\lesssim U-t$. As can be concluded from these figures, the dependence $\mu(\bar{n})$ is also rather asymmetric with respect to half-filling. This dependence is shown in Fig.~\ref{Fig3} for two values of $U$ and low temperatures. In both panels, the part corresponding to hole doping is much smaller than that to electron doping. An analogous asymmetry with respect to $\bar{n}=1$ was obtained for somewhat different parameters in \cite{Senechal} using the cluster perturbation theory. While the curve in panel (a) is smooth, in panel (b), for larger $U$, there are two nearly discontinuous regions. One of them, near $\bar{n}=1$, is related to the Mott gap. The other, near $\bar{n}\approx 0.87$, is connected with the above-mentioned destruction of the short-range AF ordering. As will be seen in the below figures, at low temperatures, for $U=5.1t$ the considered model is in the metallic state at half-filling, while for $U=8t$ it is in the insulating state. Hence at low $T$ the boundary between metallic and insulating phases is close to $U_c\approx 6t$ -- the value obtained in the cellular DMFA \cite{Park}, variation cluster approximation \cite{Balzer} and SCDT \cite{Sherman18} in the $t$-$U$ model.

\subsection{The Fermi-level peak and spin-polaron band}
In this subsection, we discuss shapes and temperature evolutions of DOSs $\rho(\omega)=N^{-1}\sum_{\bf k}A({\bf k},\omega)$ and spectral functions $A({\bf k},\omega)=-\pi^{-1}{\rm Im}G({\bf k},\omega)$ of the obtained solutions. In carrying out the analytic continuation from the imaginary to real frequency axis the maximum entropy method \cite{Press,Jarrell,Habershon} was used.

\begin{figure}[t]
\centerline{\resizebox{0.95\columnwidth}{!}{\includegraphics{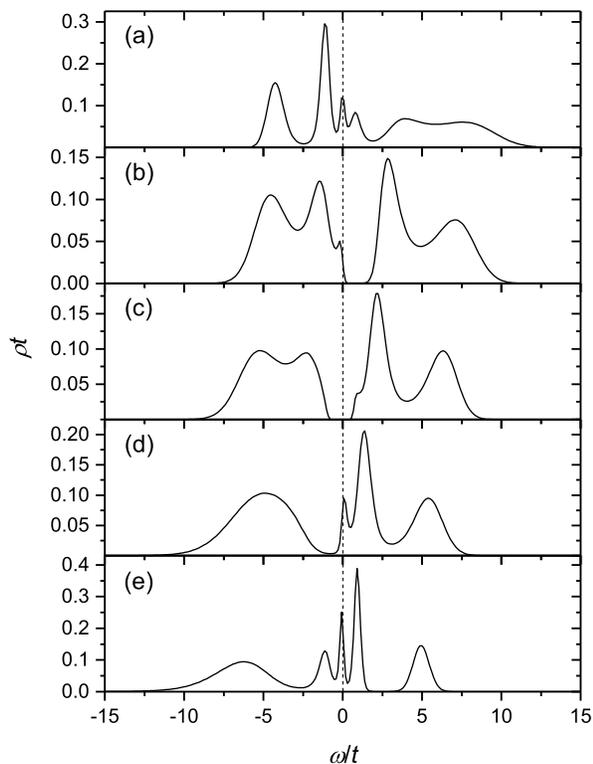}}}
\caption{The density of states for $U=8t$ and $T=0.13t$. Panels (a)-(e) correspond to $\mu=2t$ ($\bar{n}=0.91$), $\mu=3t$ ($\bar{n}=0.990$), $\mu=4t$ ($\bar{n}=0.994$), $\mu=5t$ ($\bar{n}=1.014$) and $\mu=6t$ ($\bar{n}=1.17$), respectively.} \label{Fig4}
\end{figure}

\begin{figure}[t]
\centerline{\resizebox{0.95\columnwidth}{!}{\includegraphics{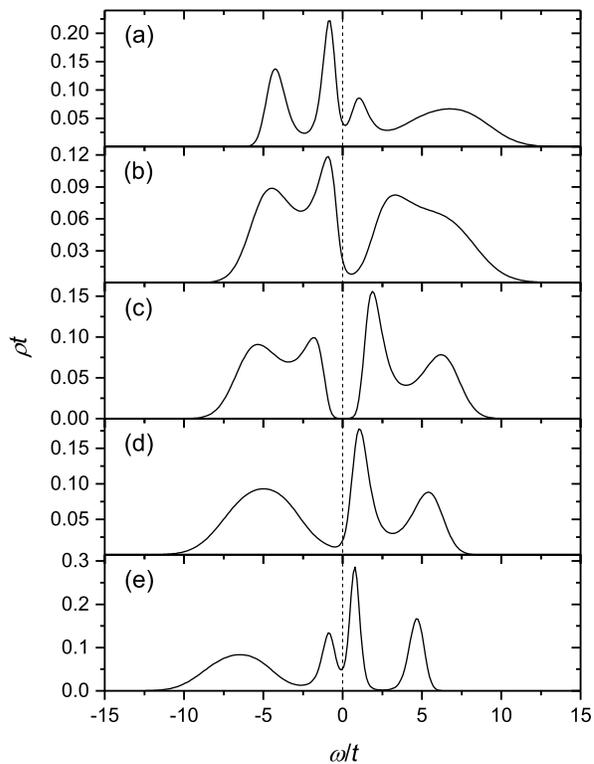}}}
\caption{The density of states for $U=8t$ and $T=0.32t$. Panels (a)-(e) correspond to $\mu=2t$ ($\bar{n}=0.92$), $\mu=3t$ ($\bar{n}=0.98$), $\mu=4t$ ($\bar{n}=1$), $\mu=5t$ ($\bar{n}=1.03$) and $\mu=6t$ ($\bar{n}=1.15$), respectively.} \label{Fig5}
\end{figure}

Figures~\ref{Fig4} and \ref{Fig5} demonstrate concentration and temperature evolutions of the DOS in the case $U=8t$. As mentioned above, for $\mu=U/2$, this case corresponds to an insulator -- the Mott gap is well seen for both temperatures in panels (c). For lo\-wer temperature, as the FL leaves the gap, there appears a sharp peak on this level. In Fig.~\ref{Fig5} co\-r\-r\-es\-pon\-ding to the slightly higher temperature, the peak disappears, though other parts of the DOS remain nearly unchanged. Panels (a) and (e) of this figure show well-developed pseudogaps around the FL. Panels (a) and (e) of Fig.~\ref{Fig4} demonstrate that with temperature decrease the peaks arises inside the pseudogaps.

\begin{figure*}[tbh]
\centerline{\resizebox{1.99\columnwidth}{!}{\includegraphics{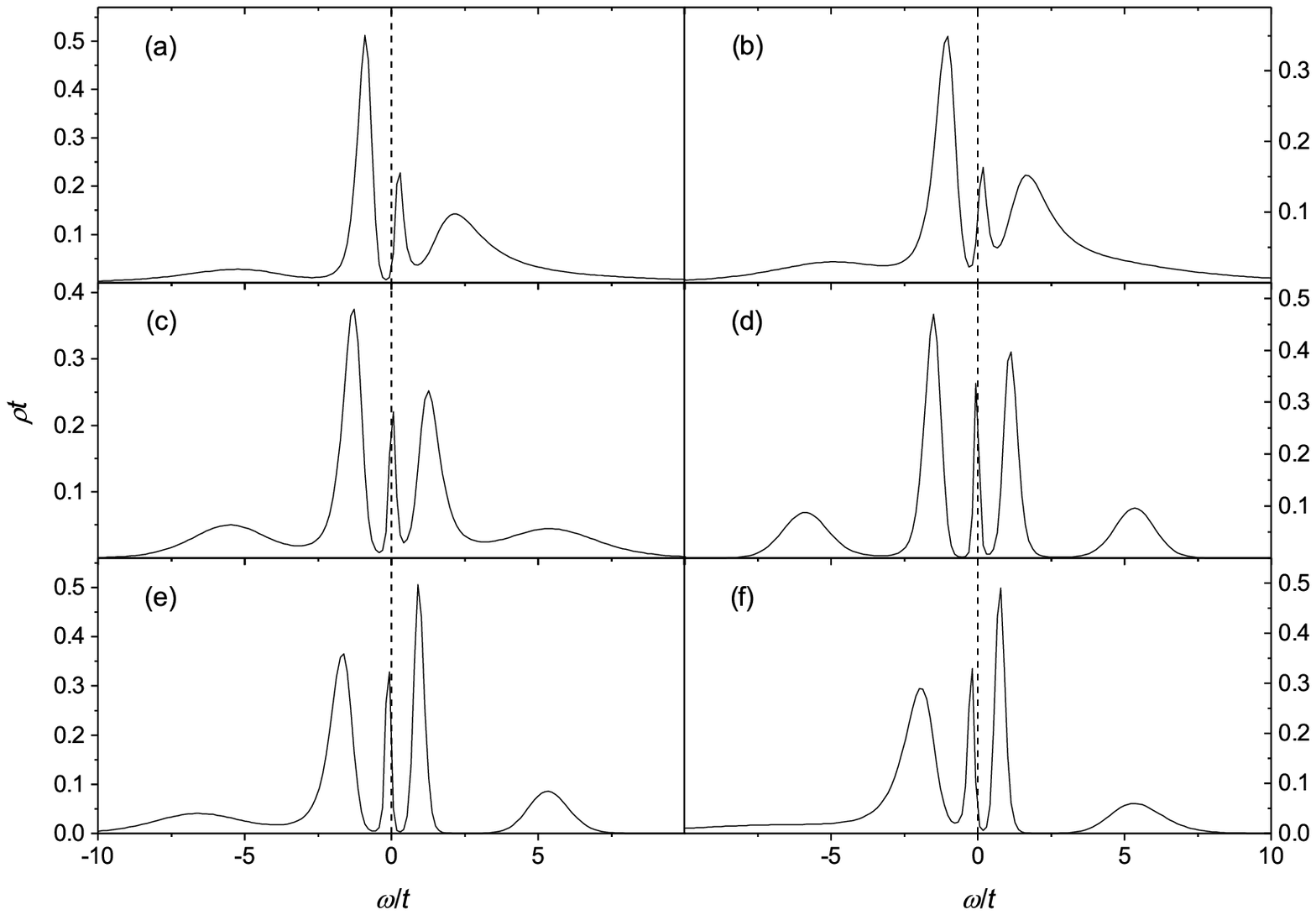}}}
\caption{The density of states for $U=5.1t$ and $T=0.12t$. Panels (a)-(f) correspond to $\mu=1.5t$ ($\bar{n}=0.95$), $\mu=2t$ ($\bar{n}=0.99$), $\mu=2.55t$ ($\bar{n}=1.05$), $\mu=3t$ ($\bar{n}=1.12$), $\mu=3.5t$ ($\bar{n}=1.2$) and $\mu=4t$ ($\bar{n}=1.28$), respectively.} \label{Fig6}
\end{figure*}

The low-temperature FL peak in the DOS is a peculiarity, which is not inherent solely to the $t$-$t'$-$t''$-$U$ model. It is observed also in the $t$-$U$ model \cite{Sherman18a}. As follows from calculations, the peak reveals itself when the maximum in $\chi({\bf k},\nu)$ at ${\bf k}={\bf Q}$ and $\nu=0$ becomes intensive, which is a manifestation of strong AF correlations. The comparison of panels (a) and (b) in Fig.~\ref{Fig2} shows that in the $t$-$t'$-$t''$-$U$ model the concentration range, in which this condition is fulfilled, is much wider. Figure~\ref{Fig6} demonstrates the FL peak in the case $U=5.1t$. For somewhat higher $T$, as in the $t$-$U$ model \cite{Sherman18}, the DOS at half-filling is inherent in a bad metal with a dip at the FL. Owing to a finite DOS at $\omega=0$, as the temperature decreases, the FL peak arises even at half-filling in this case both in the $t$-$U$ \cite{Sherman18a} and $t$-$t'$-$t''$-$U$ models. As follows from Fig.~\ref{Fig6}, in the latter model, the peak exists and is kept at the FL in the wide ranges of chemical potentials and electron concentrations.

\begin{figure}[bht]
\centerline{\resizebox{0.94\columnwidth}{!}{\includegraphics{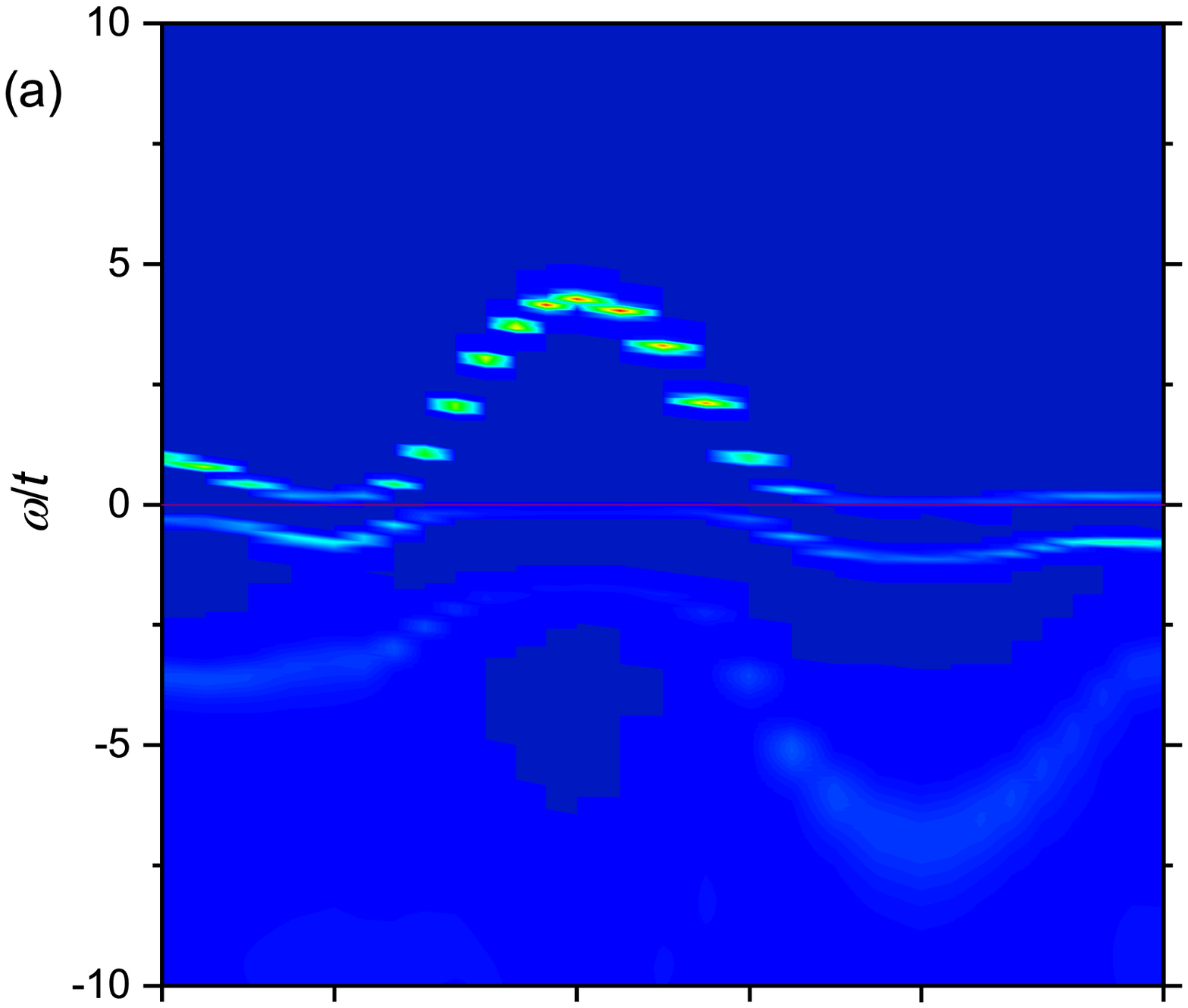}}}
\hspace{0.15em}\centerline{\resizebox{0.95\columnwidth}{!}{\includegraphics{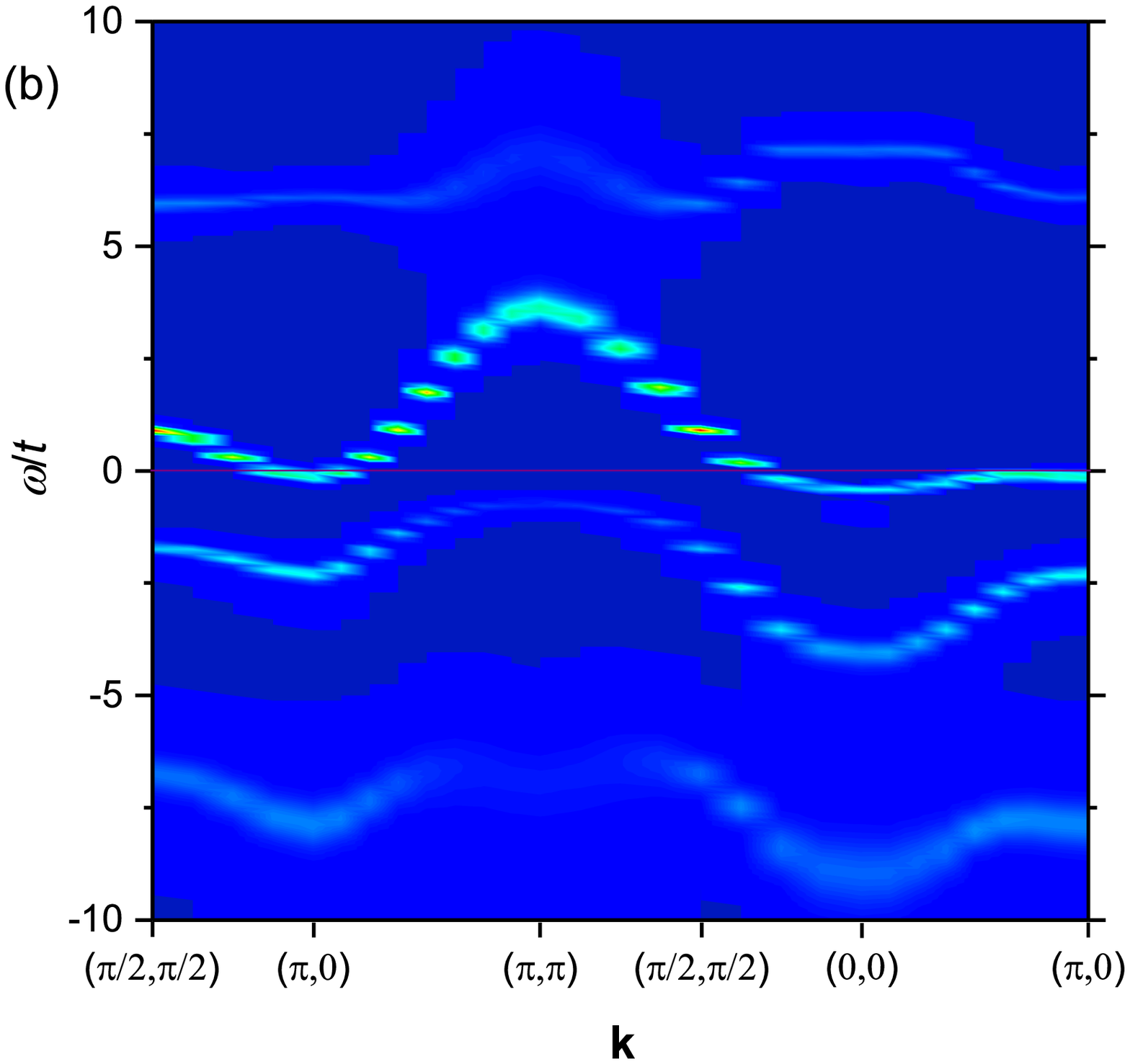}}}
\caption{The spectral function along the symmetry lines of the Brillouin zone for $U=8t$ and $\mu=6t$. Panels (a) and (b) correspond to $T=0.13t$ ($\bar{n}=1.17$) and $T=0.32t$ ($\bar{n}=1.15$), respectively. The purple line at $\omega=0$ shows the Fermi level.} \label{Fig7}
\end{figure}

What are the states responsible for the FL peaks in the above pictures? Figure~\ref{Fig7} demonstrates spectral functions along symmetry lines of the Brillouin zone for the same parameters as in Fig.~\ref{Fig4}(e) and \ref{Fig5}(e). From the comparison of two panels of Fig.~\ref{Fig7} one can see that at $T=0.13t$ new maxima arise near the FL. These maxima form a band, which hybridizes with the band crossing the FL in panel (b). As a result, there appear large regions of the Brillouin zone, in which the spectral function contains a maximum on or in the nearest vicinity of the FL. These excitations contribute to the FL peak.

Superficially the behavior of the FL peak re\-sem\-bles that of the DMFA quasiparticle peak -- both appear at low enough temperatures at the FL, at half-filling when $U$ becomes smaller than some critical value. The quasiparticle peak is a modified Kondo or Abrikosov-Suhl resonance, which is a manifestation of bound states of {\sl free} electrons and {\sl localized} spins of the AIM \cite{Georges,Hewson}. We suppose that the FL peak is also connected with bound states seen in Fig.~\ref{Fig7}. However, in our results, these are bound states of {\sl correlated} electrons and {\sl mobile} spin excitations. It counts in favor of this assumption that at half-filling and moderate $U$, when the spin-polaron band does not hybridize with other bands, its width is of the order of $J$ \cite{Sherman18a}. Besides, the band reveals itself when the maximum of $\chi({\bf k},\nu)$ at ${\bf k}={\bf Q}$ and $\nu=0$ becomes intensive, which implies a well-formed branch of spin excitations. Since the momentum dependence of the electron self-energy is rather strong in 2D \cite{Maier,Rohringer,Sherman17}, the localized picture of the DMFA based on the AIM may be inappropriate for the considered case. Flat regions of the spin-polaron band, which are responsible for the FL peak, arise due to the fact that the band is pinned to the FL, which leads to its flattening with doping. The same behavior of the spin-polaron band was observed in the $t$-$J$ model \cite{Sherman94,Sherman97}.

\begin{figure}[bht]
\centerline{\resizebox{0.95\columnwidth}{!}{\includegraphics{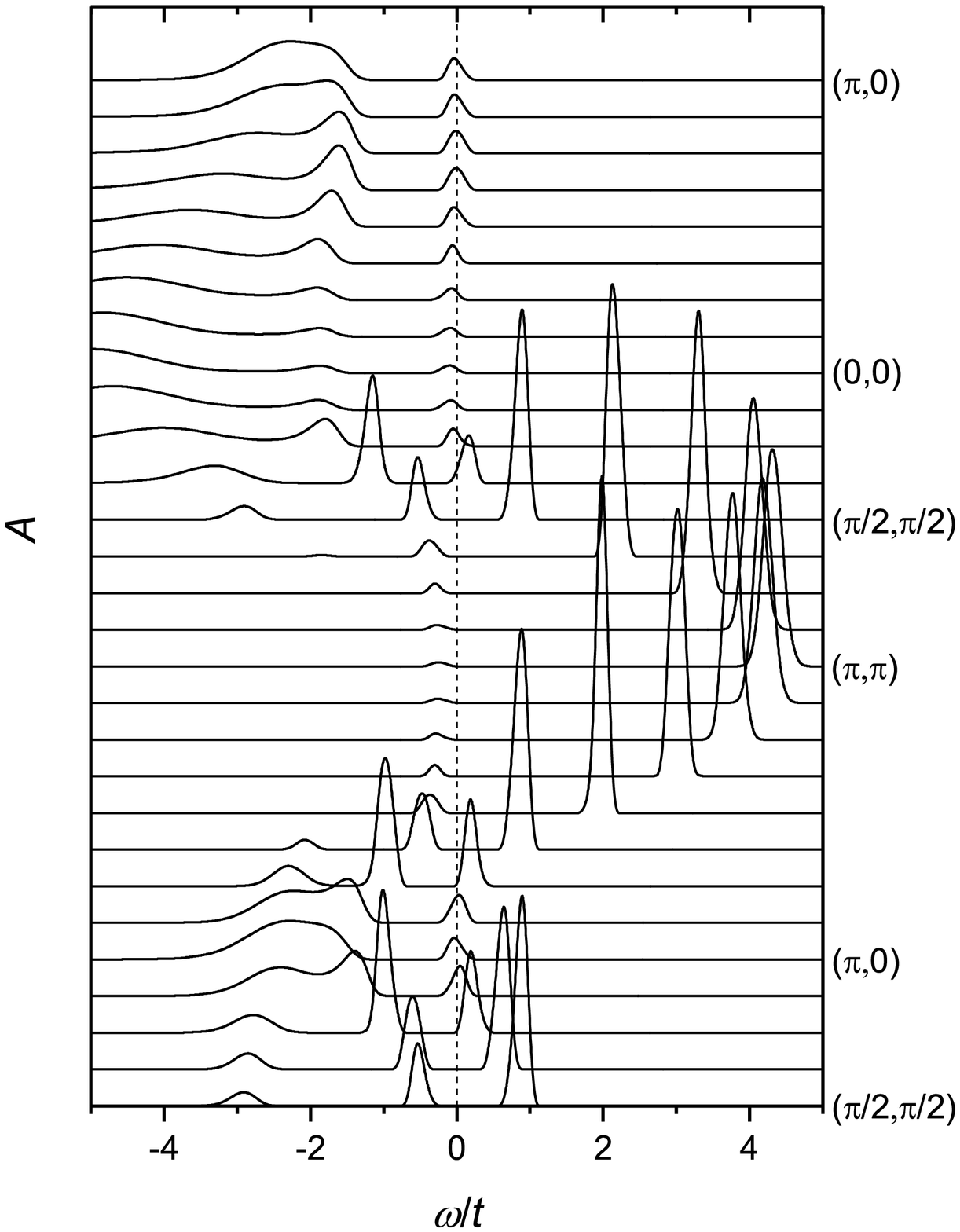}}}
\caption{The spectral function along the symmetry lines of the Brillouin zone for $U=5.1t$, $T=0.12t$ and $\mu=3.5t$ ($\bar{n}=1.2$).} \label{Fig8}
\end{figure}
Some peculiarities in photoemission spectra can supposedly be identified with the FL peak. In par\-ti\-cu\-lar, in the spectra of the electron-doped crystal Nd$_{2-x}$Ce$_x$CuO$_4$ two bands are seen near the wave vector $(\pi,0)$. One of these bands is dispersive, while another is nearly flat and located on or in the nearest vicinity of the FL \cite{Armitage01,Armitage02,Matsui05,Matsui}. As the momentum is shifted to $(\pi/2,\pi/2)$, the flat band disappears. Figure~\ref{Fig8} demonstrates calculated spectral functions in the case of electron doping. Curves near $(\pi,0)$ contain a feature with a weak dispersion at the Fermi level and a dispersive feature at a lower frequency. As the momentum approaches $(\pi/2,\pi/2)$, the dispersion of the former feature becomes stronger and it moves above the FL, disappearing from the photoemission spectrum. Thus only the latter feature remains in the spectrum for these momenta. This maximum approaches the FL as ${\bf k}\rightarrow (\pi/2,\pi/2)$. This behavior of the calculated spectra is similar to that observed in the photoemission of Nd$_{2-x}$Ce$_x$CuO$_4$. Energy scales of the experimental and calculated results are also close. If the exchange constant is set to 0.1~eV, as observed in $n$-type cuprates \cite{Armitage}, and $U=5.1t$ we find $t\approx 0.1$~eV. For such $t$ the frequency of the dispersive feature near $(\pi,0)$ in Fig.~\ref{Fig8} is close to that observed in \cite{Matsui05}. In that work, the existence of two features in photoemission spectra was explained by the hybridization of a band with its shadow arising due to the electron interaction with an AF background. Both in our calculations in Fig.~\ref{Fig7}(a), \ref{Fig8} and in results of the cluster perturbation theory \cite{Senechal} these shadow bands are very weak and, therefore, they cannot account for observed results. However, in a real crystal, intensities of these bands may be larger due to an interlayer interaction, which increases the AF correlation length or may even stabilize the long-range ordering. Dispersions of bands in our figures reflect the hybridization of a band existing at higher $T$ and the spin-polaron band. Similar structures consisting of a maximum near the FL and another maximum or shoulder at a lower frequency were observed also in photoemission of transition metal oxides (some references can be found in \cite{Georges}) and hole-doped cuprates \cite{Damascelli}. In the latter crystals, besides the foregoing, there exist other interpretations of this structure. In particular, the two spectral features may be a result of the interlayer splitting \cite{Liechtenstein}, since crystals, in which this structure was mainly studied -- Bi$_2$Sr$_2$CaCu$_2$O$_{8+\delta}$, YBa$_2$Cu$_3$O$_{7-y}$, Bi$_2$Sr$_2$Ca$_2$Cu$_3$O$_{10+\delta}$ -- are multilayer compounds.

\begin{figure}[bht]
\centerline{\resizebox{0.87\columnwidth}{!}{\includegraphics{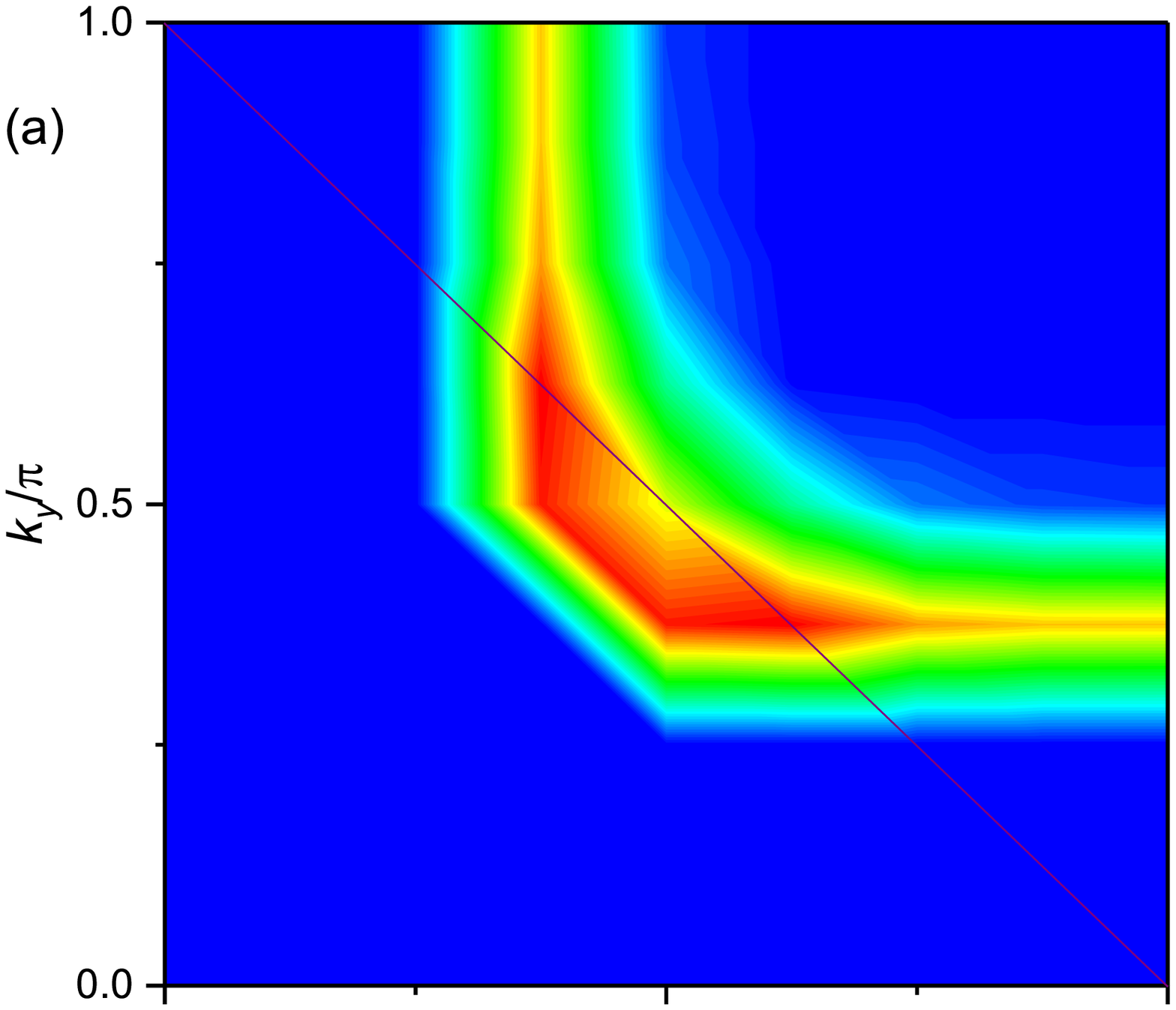}}}
\centerline{\resizebox{0.87\columnwidth}{!}{\includegraphics{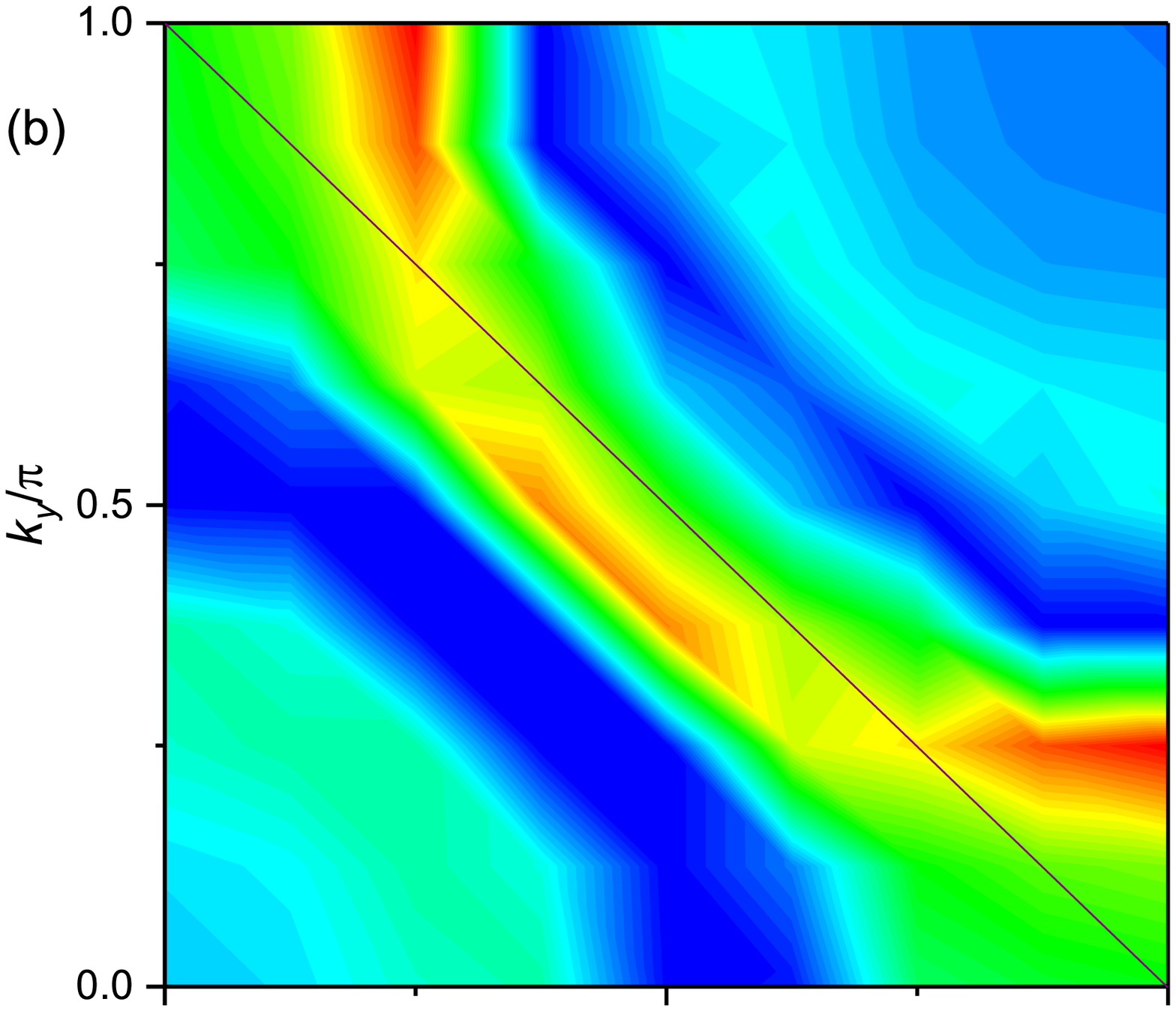}}}
\hspace{0.15em}\centerline{\resizebox{0.89\columnwidth}{!}{\includegraphics{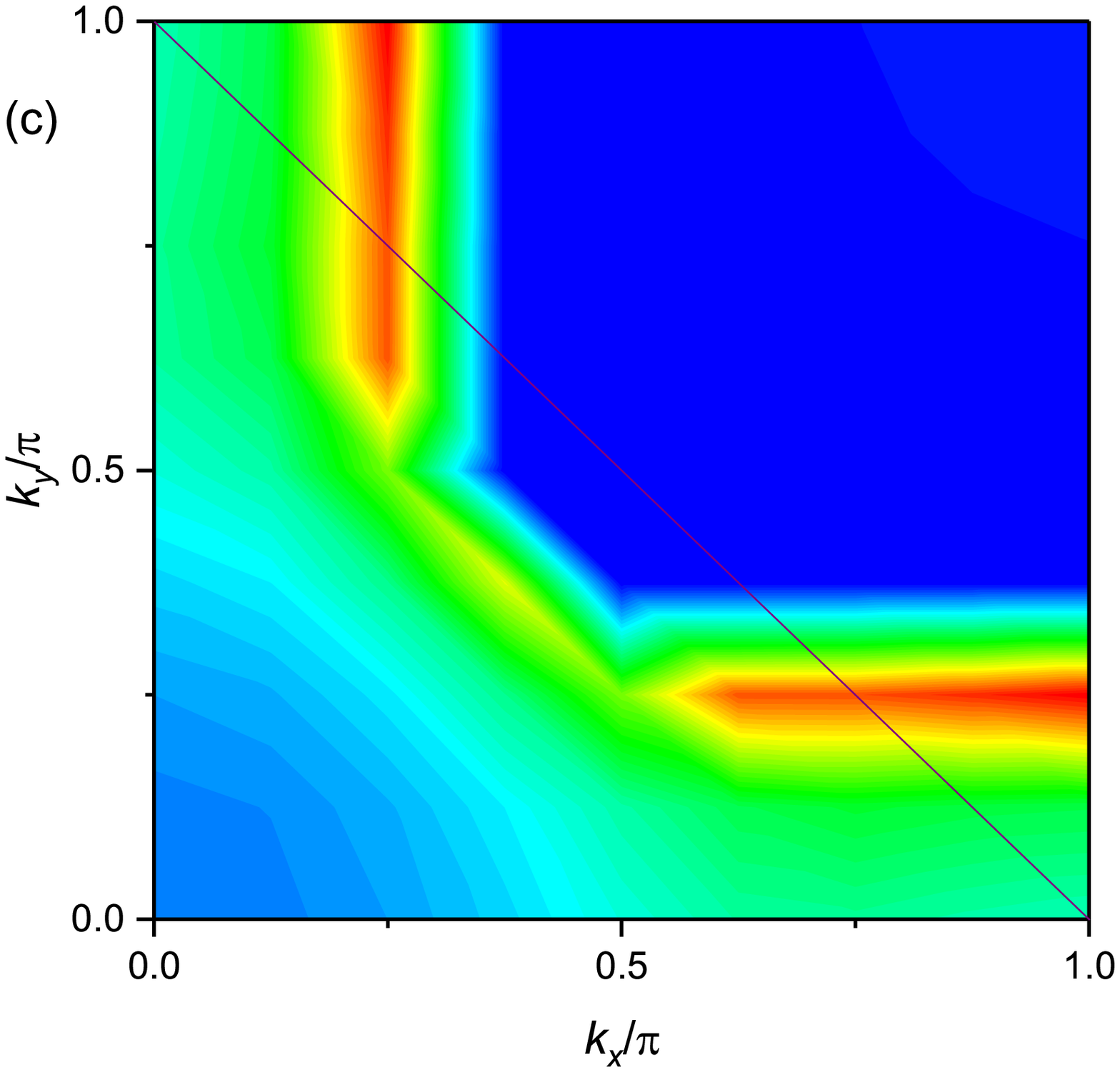}}}
\caption{The doping dependence of the Fermi contour in the first quadrant of the Brillouin zone for $U=5.1t$ and $T=0.12t$. Panels (a) to (c) correspond to $\bar{n}=0.92$, $1.05$ and $1.2$, respectively. The intensity plots were obtained by the integration of spectral functions in the frequency range from $-0.2t$ to $0.2t$. Purple lines are boundaries of the magnetic Brillouin zone.} \label{Fig9}
\end{figure}
Figure~\ref{Fig9} demonstrates the evolution of the Fermi contour when the doping varies from the hole to electron type. The intensity distribution in panel (a) is inherent in underdoped $p$-type cuprates with the Fermi arcs parallel to the boundaries of the magnetic Brillouin zone around $(\pm\pi/2,\pm\pi/2)$, $(\pm\pi/2,\mp\pi/2)$ and pseudogaps near $(\pm\pi,0)$, $(0,\pm\pi)$ \cite{Damascelli}. The most prominent features in panel (b) are intensity suppressions at hot spots -- crossing points of the Fermi contour and boundaries of the magnetic Brillouin zone. These suppressions arise due to the strong scattering of electrons by magnetic excitations with momenta near {\bf Q}. Evidently, this scattering is a manifestation of strong AF fluctuations retained for moderate doping. The ordering is gradually destroyed as the doping increases and the contour acquires the shape shown in panel (c). This evolution and contour shapes resembles that observed in Nd$_{2-x}$Ce$_x$CuO$_4$ \cite{Matsui}.

\section{Conclusion}
In this work, calculations of some spectral and mag\-ne\-tic properties of the $t$-$t'$-$t''$-$U$ fermionic 2D Hubbard model were carried out using the SCDT. Hopping constants between second and third neighbors were set to $t'=-0.3t$ and $t''=0.2$. Two values of the Hubbard repulsion -- $U=8t$ and $5.1t$ -- were considered, of which the former corresponds to an insulator and the latter to a metal at half-filling and temperatures $T\approx 0.1t$. In agreement with experiment and earlier calculations, a strong asymmetry in the behavior of AF fluctuations was found for the hole and electron doping. Using as indicators the zero-frequency susceptibility at the AF momentum {\bf Q} and the determinant of the system of linear equations, which is responsible for the strength of AF fluctuations, we found their tolerance for the electron doping and extreme sensitivity to the hole doping. For example, at $U=8t$ the above-mentioned indicators remain practically unchanged in the range of electron concentrations $1\leq\bar{n}\lesssim 1.25$, while the hole concentration $1-\bar{n}\approx 0.13$ destroys the AF ordering completely. The difference between the two sides of the phase diagram is also reflected in the dependence $\mu(\bar{n})$ and in the Fermi contours. For a moderate hole doping the contour consists of four Fermi arcs near $(\pm\pi/2,\pm\pi/2)$ and $(\pm\pi/2,\mp\pi/2)$ with pseudogaps around the points $(\pm\pi,0)$ and $(0,\pm\pi)$. For a moderate electron doping the contour is diamond-shaped with intensity suppressions in the hot spots. These suppressions originate from the scattering of electrons by intensive spin fluctuations with wave vectors from a vicinity of {\bf Q}. Shapes and evolutions of the Fermi contours are in agreement with those observed in $p$- and $n$-type cuprates.

It was found that with the temperature decrease to approximately $0.1t$ there appears a peak in the DOS at the FL. In the insulating case, the peak arises when the FL leaves the Mott gap. The FL peak is a manifestation a narrow band arising at these temperatures. In many respects, the behavior of the peak resembles that of the DMFA quasiparticle peak -- both appear at low enough temperatures at the FL, at half-filling when the Hubbard repulsion becomes smaller than some critical value. In the DMFA, the peak is a modified Kondo or Abrikosov-Suhl re\-so\-nan\-ce, which is connected with bound states of {\sl free} electrons with {\sl localized} spins. However, calculations of the electron self-energy in the 2D $t$-$U$ Hubbard model revealed its significant momentum dependence. Therefore, for the considered space dimensionality, the picture based on the AIM is in question. In this work, we identify the FL peak and the related band with bound states of {\sl correlated} electrons with {\sl mobile} spin excitations -- the notion borrowed from the $t$-$J$ model. This interpretation is supported by the width of the low-temperature band, which is of the order of the exchange constant in cases when the band does not hybridize with other bands. Besides, the peak and band arise with the appearance of a pronounced maximum in the spin susceptibility at zero frequency and the AF momentum {\bf Q}, which is an indication of a well-formed spin-excitation branch. The FL peak is not inherent solely to the $t$-$t'$-$t''$-$U$ model. In the previous work \cite{Sherman18a} it was revealed in the $t$-$U$ Hubbard model. Ranges of chemical potentials and electron concentrations, in which AF fluctuations are strong and the peak is visible, are wider in the former model. We compared calculated and photoemission spectra measured in the $n$-type crystal Nd$_{2-x}$Ce$_x$CuO$_4$ and found that some peculiarities of the experimental spectrum can be related to the FL peak.

\ack
This work was supported by the research project IUT2-27.

\end{document}